\documentclass[aps,prb,twocolumn,superscriptaddress,nobibnotes,10pt]{revtex4-2}
\usepackage{bm}
\usepackage{amsmath,amssymb}
\usepackage{graphicx}

\DeclareMathOperator{\Sp}{\mathrm{Sp}}

\DeclareMathOperator{\Ai}{Ai}

\begin{document}

\title{Dissipative quantum mechanics of Andreev bound states}
\author{Mikhail S. Kalenkov}
\affiliation{I.E.Tamm Department of Theoretical Physics, P.N.Lebedev Physical Institute, 119991 Moscow, Russia}
\author{Andrei D. Zaikin}
\affiliation{I.E.Tamm Department of Theoretical Physics, P.N.Lebedev Physical Institute, 119991 Moscow, Russia}
\affiliation{National Research University Higher School of Economics, 101000 Moscow, Russia}

\date{\today}
\begin{abstract}
We propose a microscopic scheme that allows to describe ac Josephson effect in superconducting junctions in terms of dissipative quantum dynamics of subgap Andreev bound states. This approach is particularly useful for highly transparent junctions at subgap bias voltages in which case a non-trivial combination of Landau-Zener tunneling between Andreev levels and their instability due to quasiparticle escape into continuum may play an important role. In the low bias regime, we evaluate the non-equilibrium current-phase relation of superconducting junctions at arbitrary transmissions and identify a sub-Ohmic phase-dependent dissipative contribution to the current controlled by quasiparticle dynamics near the superconducting gap edge. 
\end{abstract}
\maketitle

\section{Introduction}

Both dc and ac Josephson effects \cite{Jos,BP} in superconducting tunnel junctions are universally described by the sinusoidal dependence of the supercurrent on the phase difference between two superconducting electrodes. The situation changes as one goes beyond the tunneling limit. In equilibrium the current-phase relation (CPR) may substantially deviate from a simple sinusoidal form at sufficiently high barrier transmissions and low temperatures \cite{KO,GKI}. Further complications occur if a highly transparent Josephson junction is driven out of equilibrium. For instance, by applying a small external bias voltage $V$ one can drive the quasiparticle distribution function for such junctions far from equilibrium due to the mechanism of multiple Andreev reflections (MAR) \cite{MAR}. As a result, already at very small $V$ both $I-V$-curve \cite{Uwe} and CPR \cite{AB1,AB2} for Josephson junctions at full transmissions deviate strongly from those derived in the tunneling limit \cite{BP}.

More recently it was realized \cite{GZ1,GZ2,KZ,GZ3} that an extra important contribution to both $I-V$ curve and CPR of fully transparent junctions exists at subgap voltages. This contribution has a dissipative nature demonstrating sub-Ohmic behavior  
$\propto V^{2/3}$ for the average current \cite{GZ1,GZ2} and $\propto V^{1/3}$ for CPR \cite{KZ,GZ3} in the low voltage limit $V\ll |\Delta |/e$ 
(where $\Delta$ is the superconducting order parameter of the electrodes and $e$ is the electron charge). Furthermore, this dissipative current turns out to be periodic and strongly phase-dependent: It remains non-zero provided the Josephson phase deviates from $2\pi n$ (with $n$ integer) by not more than $\sim (e|V|/|\Delta|)^{1/3}$ and practically vanishes outside these phase intervals.

The physics behind this behavior can be understood in terms of subgap Andreev bound states which are always present inside the junction. At low voltages and provided the superconducting phase difference is far from the points $2\pi n$, the quasiparticle dynamics can be described within the Hamiltonian type-of-approach \cite{KZ26}. In this regime, the probability of excitation from Andreev bound states into the continuum is exponentially small and can be neglected. Hence, no dissipation occurs. However, as the Josephson phase approaches $2\pi n$ the quasiparticle dynamics becomes essentially non-unitary \cite{KZ} since Andreev bound states start "talking" to the continuous spectrum and eventually merge with it. As a result, quasiparticles can easily escape from their subgap states into the continuum causing a dissipative contribution to the current. 

The situation becomes more complicated for not fully transparent junctions, i.e. provided the reflection coefficient $R$ of the barrier inside the junction differs from zero. In this case the energy gap equal to $2\sqrt{R}| \Delta |$ between the two Andreev levels develops. The quasiparticle may "jump" between these subgap bound states due to the mechanism of Landau-Zener tunneling \cite{AB1} thereby causing pronounced coherent oscillations on CPR \cite{KZ26} in a certain parameter range.

In this work we will develop a detailed microscopic theory describing dissipative quantum dynamics of Andreev bound states in short superconducting junctions at subgap bias voltages. We will evaluate the dissipative contribution to the current across the junction associated with escape of quasiparticles into continuum and demonstrate an important role played by Landau-Zener tunneling in this process. Combining our analysis with the results \cite{KZ26} we formulate a complete theory of ac Josephson effect in superconducting junctions with arbitrary transmissions valid at low enough bias voltages.

The paper is organized as follows. In Sec. \ref{secmodel} we will define our model and outline basic physical picture behind the effects we are going to address. In Sec. \ref{sechamiltonian} we derive the Schr\"odinger-like equation and evaluate the wave functions for Andreev bound states within the adiabatic approximation. In Sec. \ref{secinverse} we go beyond the adiabatic approximation in order to account for instability of Andreev states at energies in the vicinity of the superconducting gap edge. Section \ref{secccurrent} is devoted to a detailed calculation of CPR and -- in particular -- of a dissipative contribution to the current across the junction. Discussion of our results and conclusions are presented in Sec. \ref{secdiscussion}. Technical details of our calculation are relegated to Appendices \ref{appA} and \ref{appB}.

\section{The model and physical picture}
\label{secmodel}
Below we will address the standard model for a single channel symmetric non-magnetic superconducting junction with transmission $D\equiv 1-R$ biased by an arbitrary time-dependent voltage $V(t)$. The phase difference $\chi (t)$ between two superconducting electrodes then reads
\begin{equation}
\chi (t)= 2e\int^tV(\tilde t)d\tilde t.
\label{phase}
\end{equation}
Note that here and below we set the Planck's constant equal to unity $\hbar=1$.

Electric current $I(t)$ can be microscopically and formally exactly evaluated within the framework of quasiclassical Eilenberger-Keldysh formalism \cite{BWBSZ} supplemented by Zaitsev boundary conditions \cite{Zaitsev} with the result \cite{KZ,KZ26}
\begin{multline}
I(t)=
ie
\Sp\Bigl\{
[W +  a ]^{-1}
[h - a h a^+]
\left([W +  a ]^{-1}\right)^+
\\
\times W^+[\chi(t)] \partial_{\chi} W[\chi(t)]
\Bigr\},
\label{T3}
\end{multline}
Here and below all the operator products are understood as convolutions in the time domain. The Andreev amplitude $a$ and the function $h$ related to the electron distribution function inside the junction are the operators with kernels $a(t-t')$ and $h(t-t')$ and their Fourier transforms
\begin{equation}
a(\varepsilon) = \dfrac{-\varepsilon + \sqrt{\varepsilon^2 - |\Delta|^2}}{|\Delta|},
\quad
h(\varepsilon) = \tanh \dfrac{\varepsilon}{2T}.
\end{equation}
The matrix $W(\chi)$ in Eq. \eqref{T3} has the form
\begin{equation}
W(\chi)  =
\begin{pmatrix}
W_d(\chi) & W_o(\chi) \\
-W_o(\chi) & W_d(\chi)
\end{pmatrix}
\label{tW}
\end{equation}
with
\begin{gather}
W_d(\chi)
=
\begin{pmatrix}
\sqrt{D} \cos(\chi/2) & \sqrt{R} \\
\sqrt{R}  & -\sqrt{D} \cos(\chi/2)
\end{pmatrix},
\\
W_o(\chi) = \sqrt{D} \sin(\chi/2)
\begin{pmatrix}
1 & 0 \\
0 & 1
\end{pmatrix}.
\end{gather}

\begin{figure}
\begin{center}
\includegraphics[width=90mm]{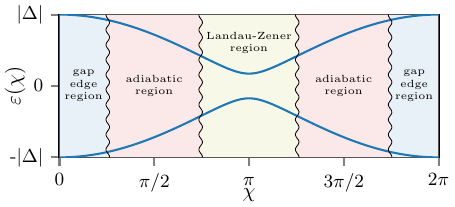}
\end{center}
\caption{The energy of Andreev bound states as a function of the superconducting phase difference $\chi$. Adiabatic dynamics and the process of Landau-Zener tunneling contribute to the Josephson current $I_J[\chi (t)]$. Superconducting gap edge regions close to $\chi =2\pi n$ yield an extra dissipative contribution to the current $I_{\rm diss}[\chi (t)]$.}
\label{ABS-paper-fig}
\end{figure}

Before we dwell into a detailed calculation let us pause and turn to the basic physical picture we are going to describe. This picture is essentially based on the presence of the two Andreev bound states inside the junction with phase-dependent energies $\pm \varepsilon_A(\chi)$, where
\begin{equation}
\varepsilon_A(\chi) = |\Delta|\sqrt{1 - D \sin^2(\chi/2)}.
\label{Andreev}
\end{equation}
These subgap bound states are schematically depicted in Fig. \ref{ABS-paper-fig}. By applying a (small) bias voltage $V$ one adiabatically sweeps the Josephson phase $\chi (t)$ \eqref{phase} and increases the energy of a quasiparticle inside the junction driving it along the lower Andreev level from the state with $\varepsilon =-|\Delta |$ upwards. As the quasiparticle reaches the anticrossing region $\chi \approx \pi$ it can either suffer Landau-Zener tunneling and jump to the higher Andreev level or remain in the lower Andreev state. Further system dynamics with $\chi > \pi$ involves both Andreev levels. Careful quantum mechanical analysis of adiabatic evolution of our system together with the process of Landau-Zener tunneling \cite{KZ26} allows to evaluate a non-dissipative Josephson current as a functional of the time-dependent superconducting phase $I_J[\chi (t)]$.

This analysis becomes insufficient in the vicinity of the phase values $\chi =2\pi n$ where discrete Andreev levels approach the continuum of states above the superconducting gap $|\Delta |$. In these regions the bound states become unstable because a quasiparticle can easily escape into the continuum thereby causing an extra -- periodic in $\chi$ -- dissipative contribution to the current $I_{\rm diss}[\chi (t)]$. As a result, the total phase-dependent current across the junction at subgap bias voltages is given by the sum of two functionals of $\chi(t)$:
\begin{equation}
I[\chi (t)]=I_J[\chi (t)]+I_{\rm diss}[\chi (t)].
\label{sum}
\end{equation}

Below we will evaluate the dissipative current  $I_{\rm diss}[\chi (t)]$ and establish a non-equilibrium CPR \eqref{sum} for Josephson junctions with arbitrary transmission values $D$ exposed to a small bias voltage $V$.

\section{Effective Hamiltonian and wave functions}
\label{sechamiltonian}
In what follows we will assume that the phase variable $\chi(t)$ increases monotonously with time (implying that $V(t)>0$) and define the time points
\begin{gather}
\label{plus}
\chi(t_{0, k, c}) = 4\pi k + \pi c - \pi,
\\
\chi(t_{\pi, k, c}) = 4\pi k + \pi c,
\\
\chi(t_{2\pi, k, c}) = 4\pi k + \pi c+ \pi,
\label{minus}
\end{gather}
where $k$ is an integer and $c = \pm 1$. In this way the entire time axis $-\infty < t < \infty$ is decomposed into a sequence of $2\pi$ intervals $t_{0, k, c} < t < t_{2\pi, k, c}$. These intervals do not overlap with each other and cover the whole time axis without gaps. Adjacent time intervals match at their boundaries according to the relation
\begin{equation}
t_{2\pi, k, c} = t_{0, k + 1/2 + c/2, -c}.
\label{t0t2pi}
\end{equation}
Note that each of the above phase values and time points are parametrized by two -- rather than just one -- different numbers $k$ and $c$. The meaning of this choice will be clear below.

Provided the phase difference $\chi(t)$ is not too close to the points $\chi = 2\pi n$ (with $n=0,\pm1,\pm2,...$), the junction dynamics at low voltages $eV \ll |\Delta |$ can be accounted for by means of the Hamiltonian-like approach \cite{KZ26}. To this end we will make use of the following exact operator identities
\begin{gather}
\begin{split}
(W &+  a)^{-1}
=
-\dfrac{|\Delta|}{2}
\begin{pmatrix}
(\varepsilon - \mathcal{H}_L)^{-1} & 0 \\
0 & (\varepsilon - \mathcal{H}_L)^{-1}
\end{pmatrix}
\\&\times
\Biggl[
\begin{pmatrix}
W_o a^{-1} W_o^{-1} & 0
\\
0 & W_o a^{-1} W_o^{-1}
\end{pmatrix}
W^+ + 1
\Biggr],
\end{split}
\label{inv2_L}
\\
\begin{split}
(W &+  a)^{-1}
=-\dfrac{|\Delta|}{2}
\Biggl[
W^+
\begin{pmatrix}
W_o^{-1} a^{-1} W_o & 0
\\
0 & W_o^{-1} a^{-1} W_o
\end{pmatrix}
\\&+ 1
\Biggr]
\begin{pmatrix}
(\varepsilon - \mathcal{H}_R)^{-1} & 0 \\
0 & (\varepsilon - \mathcal{H}_R)^{-1}
\end{pmatrix} ,
\end{split}
\label{inv2_R}
\end{gather}
where we introduced the "left" and "right" effective Hamiltonians
\begin{gather}
\mathcal{H}_L =
\dfrac{|\Delta|}{2}
\Bigl[
W_d
+
W_o a^{-1}  W_o^{-1} W_d a
+
W_o a^{-1}  W_o^{-1}
-
a^{-1}
\Bigr],
\label{Heff_L}
\\
\mathcal{H}_R =
\dfrac{|\Delta|}{2}
\Bigl[
W_d
+
a W_d W_o^{-1} a^{-1}  W_o
+
W_o^{-1} a^{-1}  W_o
-
a^{-1}
\Bigr].
\label{Heff_R}
\end{gather}
Note that only one of the above identities \eqref{inv2_L} or \eqref{inv2_R} and respectively one of the Hamiltonians \eqref{Heff_L} or \eqref{Heff_R} would already be sufficient for our consideration. Technically, however, it would be more convenient to employ all four equations \eqref{inv2_L}-\eqref{Heff_R} for the sake of compactness of our further analysis.

The effective Hamiltonians $\mathcal{H}_{L,R}$ are in general nonlocal in time (and, hence, non-Hermitian) integral operators. In the limit of low voltages, however, they can be reduced to conventional local in time operators because the matrices $W_{d,o}$ remain slowly varying functions of time in this limit. As a result, they can be approximately commuted with the operators $a^{-1}$ nonlocal only on a short time scale of order $1/|\Delta|$. This is achieved by means of the following approximate interchange rule
\begin{equation}
F(\varepsilon) f(t) \approx
f(t) F(\varepsilon) + i f'(t) F'(\varepsilon).
\label{interRL}
\end{equation}
For compactness, here we employ a symbolic notation: The product $F(\varepsilon) f(t)$ implies that the integral operator with the kernel $F(t-t')$ (and Fourier transform $F(\varepsilon)$) acts on the function $f(t)$, and similarly for $f(t)F(\varepsilon)$.

After a straightforward algebra we obtain
\begin{equation}
\mathcal{H}_{L,R}(t) =
\mathcal{H}_0[\chi(t)]
\pm
\dfrac{i\dot \chi(t)}{4}
\cot [\chi(t)/2]
\mp
\dfrac{\dot \chi(t)}{4}
c \hat p_3,
\label{Heff3L}
\end{equation}
Here and below the upper (lower) sign corresponds to the index $L$ ($R$), 
the time is restricted to the interval $t_{0,k,c}<t<t_{2\pi,k,c}$, $\hat p_{1,2,3}$ are Pauli matrices, and the matrix Hamiltonian $\mathcal{H}_0$ is defined as
\begin{equation}
\mathcal{H}_0(\chi) =
|\Delta|
\begin{pmatrix}
\sqrt{D} \cos (\chi/2) & \sqrt{R}  \\
\sqrt{R} & -\sqrt{D} \cos (\chi/2)
\end{pmatrix}.
\label{H0}
\end{equation}

Our next step is to introduce the corresponding Schr\"odinger-like equations
\begin{equation}
i\dfrac{\partial \overline{\Psi}_{L,R}}{\partial t}
= \mathcal{H}_{L,R} \overline{\Psi}_{L,R}.
\label{PsiLRShrod}
\end{equation}
At this point it is appropriate to stress that, while the nonlocal effective Hamiltonians \eqref{Heff_L}, \eqref{Heff_R} (and the corresponding Eqs. \eqref{inv2_L}, \eqref{inv2_R}) are exact, the local in time Hamiltonians \eqref{Heff3L} are obtained within the low voltage approximation and, hence, should be treated with sufficient care. For instance, we observe that these Hamiltonians contain terms diverging at $\chi(t) = 2\pi n$ and thereby signaling the failure of the above approximation in the vicinity of these points.

Employing the substitutions
\begin{gather}
\overline{\Psi}_L(t) = \sqrt{|\sin[\chi(t)/2]|} \tilde \Psi_L(t),
\\
\overline{\Psi}_R(t) = \dfrac{\tilde \Psi_R(t)}{\sqrt{|\sin[\chi(t)/2]|}} ,
\end{gather}
one may rewrite the Schr\"odinger-like equations \eqref{PsiLRShrod}  as
\begin{equation}
i\dfrac{\partial \tilde \Psi_{L,R}}{\partial t}
= \tilde{\mathcal{H}}_{L,R} \tilde \Psi_{L,R}
\label{shrodLR}
\end{equation}
with Hermitian effective Hamiltonians
\begin{equation}
\tilde{\mathcal{H}}_{L,R}
=\mathcal{H}_0 \mp
\dfrac{\dot \chi(t)}{4} c \hat p_3 .
\label{ham}
\end{equation}
Here it is important to retain the last term proportional to $\dot \chi(t)$ which -- being small as compared to $\mathcal{H}_0$ at low voltages --  may nevertheless yield substantial modifications of the "wave functions" $\tilde \Psi_{L,R}$ in the long time limit.

Note that these two functions are linked to each other by means of the transformation 
\begin{equation}
\tilde \Psi_R(t) = bQ_c[\chi(t)] \tilde \Psi_L(t),
\label{trans}
\end{equation}
where $b$ is an arbitrary complex number obeying the condition $|b|=1$ and 
\begin{equation}
Q_c(\chi) =
\begin{pmatrix}
\sqrt{D} e^{-ic\chi/2} & \sqrt{R} \\
\sqrt{R} & -\sqrt{D} e^{ic\chi/2}
\end{pmatrix}.
\end{equation}
The relation \eqref{trans} makes it sufficient to resolve only one of the equations ~\eqref{shrodLR}.  Away from the points $t_{\pi,k,c}$ the corresponding solution can be found within the adiabatic approximation which yields \cite{KZ26}
\begin{multline}
\psi_{k, c; L(R), \mu}(t) =
\dfrac{\varphi_{\mu}[\chi(t)]}{
\sqrt{a(\pm\mu \varepsilon_A[\chi(t)])}}\theta(t - t_{0, k, c}) \theta(t_{2\pi, k, c}-t)\\\times
\exp\left(
-i \mu \int_{t_{\pi, k, c}}^t\varepsilon_A[\chi(\tilde t)] d\tilde t\right),
\label{adiabatic}
\end{multline}
where the wave functions with indices $(k,c)$ are restricted to the time interval $t_{0, k, c} < t < t_{2\pi, k, c}$. The function  $\varepsilon_A(\chi)$ is defined in Eq. \eqref{Andreev} and \cite{KZ26}
\begin{equation}
\varphi_{+}(\chi) = \varphi(\chi),
\quad
\varphi_{-}(\chi) = i\hat p_2 \varphi(\chi)
\end{equation}
with
\begin{equation}
\varphi(\chi)
=
\dfrac{
\begin{pmatrix}
\sqrt{R} |\Delta| \\
\varepsilon_A(\chi) - \sqrt{D} |\Delta | \cos (\chi/2)
\end{pmatrix}}{
\sqrt{2 \varepsilon_A(\chi) \left[ \varepsilon_A(\chi) - \sqrt{D} |\Delta | \cos (\chi/2)\right]}
}.
\label{vector}
\end{equation}

We now introduce a set of orthonormal solutions $\tilde \Psi_{k, c; X,\mu}(t)$ ($X=L,R$) of the Eq. \eqref{shrodLR}, which coincide with the corresponding adiabatic wave functions $\psi_{k, c; X, \mu}(t)$ at $t \to t_{0,k,c}$. In addition to the orthonormality condition
\begin{equation}
\tilde \Psi^+_{k, c; X,\mu}(t)  \tilde \Psi_{k, c; X,\nu}(t) = \delta_{\mu,\nu},
\end{equation}
these solutions obey the completeness relation
\begin{equation}
\sum_{\mu} \tilde \Psi_{k, c; X,\mu}(t) \tilde \Psi^+_{k, c; X,\mu}(t)   = 1
\end{equation}
as well as the symmetry relation
\begin{equation}
\tilde \Psi^*_{k, c; X,\mu}(t) =  \mu \hat p_2 \tilde \Psi_{k, c; X,-\mu}(t).
\label{psi*psi}
\end{equation}
The same relations also hold for the corresponding adiabatic wave functions $\psi$.

Employing this basis, one can write explicit expressions for the integral kernels of the inverse operators
$(\varepsilon - \mathcal{H}_{L,R})^{-1}$. We obtain
\begin{multline}
(\varepsilon - \mathcal{H}_X)^{-1}
=
-i \theta(t - t')
\Biggl[\dfrac{|\sin [\chi(t)/2]|}{|\sin [\chi(t')/2]|}\Biggr]^{\pm1/2}
\\\times
\sum_{k,c}
\sum_{\mu =\pm}
\tilde \Psi_{k, c; X,\mu}(t)
\tilde \Psi^+_{k, c; X,\mu}(t'),
\label{invH_L}
\end{multline}
where, as before, the upper (lower) sign in front of the power value 1/2 corresponds to $X=L$ ($X=R$).

It is convenient to represent the wave functions $\tilde \Psi_{k,c;L(R),\mu}$ as linear combinations of the adiabatic states $\psi_{k,c;L(R),\nu}$ as
\begin{equation}
\tilde \Psi_{k, c; X,\mu}(t) =
\sum_{\nu}
\mathcal{S}_{k,c;\nu,\mu}(t)\psi_{k, c; X,\nu}(t),  
\label{PsiSL}
\end{equation}
where $X=L,R$ and $\mathcal{S}_{k,c;\nu\mu}(t)$ are the elements of a unitary scattering matrix $\mathcal{S}_{k,c}(t)$.
One can first introduce the relation \eqref{PsiSL} only for the "left" functions $X=L$ and then verify that the "right" ones $X=R$ also obey it
due to the transformations
\begin{equation}
\tilde \Psi_{k, c; R, \mu}(t)
=
i Q_c
\tilde \Psi_{k, c; L, \mu}(t)
\label{PsiQPsi}
\end{equation}
and 
\begin{equation}
\psi_{k, c; R, \mu}(t) = i Q_c \psi_{k, c; L, \mu}(t),
\label{psiQpsi}
\end{equation}
which are in accordance with  Eq. \eqref{trans} with $b=i$.

Making use of the symmetry relation \eqref{psi*psi}, we arrive at the following constraint
\begin{equation}
\mathcal{S}_{k,c;\nu,\mu}(t) = \mu\nu\mathcal{S}^*_{k,c;-\nu,-\mu}(t),
\label{Srel}
\end{equation}
which allows one to parametrize the matrix $\mathcal{S}_{k,c}$ in the form
\begin{equation}
\mathcal{S}_{k,c}
=
\begin{pmatrix}
r_{k,c}(t) & d_{k,c}(t) \\
-d_{k,c}^*(t) & r_{k,c}^*(t)
\end{pmatrix},
\label{Smatrix}
\end{equation}
where the complex amplitudes $r_{k,c}(t)$ and $d_{k,c}(t)$ satisfy the unitarity condition $|r_{k,c}(t)|^2 + |d_{k,c}(t)|^2 = 1$.

\section{Inverse operator}
\label{secinverse}
Now we have all the necessary ingredients in order to evaluate the inverse operator $(W +  a)^{-1}$. Substituting Eq.~\eqref{invH_L} with $X=L$ into Eq.~\eqref{inv2_L}
(or, equivalently,  Eq.~\eqref{invH_L} with $X=R$ into Eq.~\eqref{inv2_R}), we obtain
\begin{multline}
(W +  a)^{-1}
=
\dfrac{\sqrt{D}|\Delta|}{2}
\theta(t - t')
\sum_{k,c}
\sum_{\mu =\pm}
\Psi_{k, c; L,\mu}(t)
\\\times
\begin{pmatrix}
i & -c \\
c & i
\end{pmatrix}
\Psi^+_{k, c; R,\mu}(t'),
\label{invW}
\end{multline}
where we introduced the notation
\begin{equation}
\Psi_{k, c; X,\mu}(t) = \sqrt{ |\sin [\chi(t)/2]|} \tilde \Psi_{k, c; X,\mu}(t),
\quad
X = L,R,
\label{def}
\end{equation}
and the product $\Psi_{k,c;L,\mu}(t) M \Psi^+_{k,c;R,\mu}(t')$  in Eq. \eqref{invW} denotes the block matrix with $(i,j)$ block $m_{ij}\Psi_{k,c;L,\mu}(t)\Psi^+_{k,c;R,\mu}(t')$, with $m_{ij}$ are the entries of $M$.

The result \eqref{invW} demonstrates that the form of the inverse operator $(W +  a)^{-1}$ turns out to be somewhat different for adjacent time intervals with $c=\pm 1$, thereby explaining our choice \eqref{plus}-\eqref{minus} involving two numbers $k$ and $c$.

Although Eq. \eqref{invW} formally involves a summation over all values of $k$ and $c$, in practice only one term contributes for any given pair of time arguments $t$ and $t'$. Indeed, the wave functions $\Psi_{k,c;X,\mu}(t)$ are localized within the corresponding $2\pi$ phase interval and vanish outside it. Since the intervals labeled by different pairs $(k,c)$ do not overlap, only one pair $(k,c)$ yields a nonzero contribution to Eq. \eqref{invW}.

Equation \eqref{invW} combined with the definition \eqref{def} remains valid provided the time moments $t$ and $t'$ are well separated, i.e. $|t - t'| \gg 1/|\Delta|$, and, on top of that, provided the phase values $\chi(t)$ and $\chi(t')$ are not too close to the points $\chi = 2\pi n$.
In order to repair this problem and to evaluate the current in the vicinity these points it is necessary to generalize the expression for the inverse operator $(W +  a)^{-1}$ in a way to make it applicable for the entire time axis. This task will be accomplished below.

Before proceeding further with our analysis let us make an important observation. Substituting the inverse operator \eqref{invW} into Eq.~\eqref{T3}, one obtains the following combination
\begin{multline}
k_{\mu\nu} (t) =
\int_{-\infty}^t d t_1 \int_{-\infty}^t d t_2
\Psi^+_{k, c; R,\mu}(t_1)
\\\times
K(t_1 - t_2)
\Psi_{k, c; R,\nu}(t_2),
\label{K}
\end{multline}
where the Fourier transform of the kernel $K(t_1 - t_2)$ reads
\begin{equation}
K_{\varepsilon} =  h (\varepsilon) [1 - |a(\varepsilon)|^2].
\end{equation}
At subgap energies $|\varepsilon| < |\Delta|$ one has $|a(\varepsilon)|^2 \equiv 1$ and, hence, $K_{\varepsilon}=0$. As a consequence, the operator $K$ annihilates any function with the Fourier components of the form $e^{-iEt}$ with subgap energies $|E|<|\Delta|$.
In Eq.~\eqref{K}, the operator $K$ acts on the wave function $\Psi_{k, c; R,\nu}(t)$, which Fourier transform is concentrated in the immediate vicinity of the energies $\varepsilon = \nu \varepsilon_A[\chi(t)]$. Away from the points $\chi = 2\pi n$ these energies are confined to the subgap domain and, hence, the corresponding contribution vanishes. 

It follows immediately that the dominant contribution to the integrals over $t_1$ and $t_2$ in Eq. \eqref{K} is determined by the domain with both time variables being localized near the lower endpoint $t_{0,k,c}$ of the corresponding $2\pi$ interval, i.e. where the Andreev levels merge with the continuum and the condition $|a(\varepsilon)|^2=1$ no longer holds. On top of that, provided the observation time $t$ approaches the upper endpoint $t_{2\pi,k,c}$ of the same interval, an extra contribution arises with both $t_1$ and $t_2$ being localized near $t_{2\pi,k,c}$.
In all these cases the wave functions $\tilde\Psi_{k,c;R,\mu}$ can no longer be described by the Schr\"odinger-like equations \eqref{shrodLR} with the local in time effective Hamiltonians \eqref{ham}.

In order to analyze the behavior of the wave functions and the inverse operator in the vicinity of the phase values $\chi = 2\pi n$ it would be convenient to perform the following matrix rotation
\begin{equation}
\tilde W
=
U^+ W U
=
\begin{pmatrix}
W_{-1} & 0 \\ 0 & W_{1}
\end{pmatrix}
,
\quad
U=
\dfrac{1}{\sqrt{2}}
\begin{pmatrix}
1 & i \\ i & 1
\end{pmatrix},
\end{equation}
where we defined 
\begin{equation}
W_{\pm 1} = W_d \mp i W_o.
\label{Wpm}
\end{equation}
 In this rotated basis the inverse operator $(\tilde W + a)^{-1}$ takes the form
\begin{multline}
(\tilde W +  a)^{-1}
=
\dfrac{i\sqrt{D}|\Delta|}{2}
\theta(t - t')
\\\times
\sum_{k,c}
\sum_{\mu =\pm}
\Psi_{k, c; L,\mu}(t)
\begin{pmatrix}
1-c & 0 \\
0 & 1+c
\end{pmatrix}
\Psi^+_{k, c; R,\mu}(t'),
\label{invW2}
\end{multline}
where we employ the same compact notation for the matrix product $\Psi_{k,c;L,\mu}(t) M \Psi^+_{k,c;R,\mu}(t')$ as in Eq.~\eqref{invW}.
Alternatively Eq. \eqref{invW2} can be written as
\begin{multline}
(W_{c} +  a)^{-1}
=
i\sqrt{D}|\Delta|
\theta(t - t')
\\\times
\sum_{k}
\sum_{\mu =\pm}
\Psi_{k, c; L,\mu}(t)
\Psi^+_{k, c; R,\mu}(t').
\label{invW3}
\end{multline}

Thus far we only performed identical transformations of the inverse operator \eqref{invW} and, hence, the validity range of Eq.~\eqref{invW3} combined with Eq. \eqref{def} should coincide with that of Eq.  \eqref{invW}. Hence, these equations {\it do not} apply provided the time variables $t$ and $t'$ are such that $\chi(t)$ or $\chi(t')$ approach the vicinity of the points $2\pi n$.

Now we make an important step. We keep the same form of the inverse operator \eqref{invW3} for {\it all} values of $\chi(t)$ and $\chi(t')$ (but still assuming $|t-t'| |\Delta|\gg 1$) and abandon the definition of the wave functions \eqref{def}. In this case the wave functions should be determined from the integral equation which follows directly from Eq.~\eqref{invW3} and can symbolically be written in the form
\begin{equation}
\Psi^+_{k,c;R,\mu}(W_{c}+a)=0.
\label{PsiRW}
\end{equation}
It is easy to observe that Eq.~\eqref{invW3} combined with Eq. \eqref {PsiRW} remains a valid representation of the inverse operator for all phase values including $\chi(t)$ and $\chi(t')$ close or equal to $2\pi n$.
This observation can be verified, e.g., by multiplying Eq.~\eqref{invW3} by the operator $(W_{c}+a)$ from the right and making use of Eq.~\eqref{PsiRW}. Sufficiently far from the phase values $\chi(t) = 2\pi n$ and at low voltages  Eq.~\eqref{PsiRW} reduces to
the Schr\"odinger-like equation \eqref{shrodLR} with the Hamiltonian \eqref{ham}.  On the other hand, close to these phase values Eq.~\eqref{PsiRW} deviates from Eq.  \eqref{shrodLR} providing an appropriate description of the wave function $\Psi_{k, c; R,\mu}$ in the regions where the Schr\"odinger-like description is no longer sufficient.

To conclude, it order to correctly evaluate CPR for our superconducting junction at subgap bias voltages and all values of the Josephson phase $\chi (t)$ one needs to first recover the proper wave functions by solving the integral equation ~\eqref{PsiRW} and then substitute the corresponding expressions into Eq. ~\eqref{invW3} for the inverse operator $(W_{c} +  a)^{-1}$. Technically this program is carried out in Appendices \ref{appA} and \ref{appB}. Below we will make use of the corresponding results and derive the electric current $I [\chi (t)]$.

\section{Electric current}
\label{secccurrent}

Let us begin our analysis by evaluating the current for the phase values being sufficiently far from the points $\chi=2\pi n$, i.e. the time variable $t$ belonging to the interval $(t_{0,k,c},t_{2\pi,k,c})$ should not approach its endpoints. In this regime we substitute the inverse operator in the form \eqref{invW} into Eq.~\eqref{T3}. As discussed above, the integration over the internal time variable $t'$ is localized near the lower endpoint $t_{0,k,c}$ of the corresponding interval. Hence, for the ``right'' wave functions entering Eq. \eqref{invW} one should employ the asymptotic expressions derived in Appendix \ref{appA}. The resulting $t'$ integration is encoded in the matrix elements $k_{\mu\nu}(t)$ defined in Eq. \eqref{K}, which can be evaluated explicitly with the aid of Eqs. \eqref{tpsip+} and \eqref{tpsim+}. For this purpose, it is convenient to employ the Fourier representation of the functions $f_{R,\pm}(\tau)$ \eqref{fmusol+} rather than their explicit expressions in terms of the Airy functions. After a straightforward integration we obtain
\begin{equation}
k_{\mu\nu} (t) = \dfrac{h_{\mu}}{\sqrt{D}|\Delta|}\delta_{\mu,\nu},
\quad h_{\pm} = \tanh[\pm |\Delta|/(2T)]
\end{equation}
and arrive at a general expression for the current  in the form
\begin{equation}
I(t)=
\dfrac{1}{2}
\sum_{k,c;\mu}
h_{\mu}
\Psi^+_{k,c;L,\mu} (t)\hat{\mathcal{J}}_c[\chi (t)] \Psi_{k,c;L,\mu} (t),
\label{T8}
\end{equation}
where we introduced the current operator
\begin{equation}
\hat{\mathcal{J}}_c (\chi)=
e D |\Delta|
c
\begin{pmatrix}
\sqrt{D} & \sqrt{R} e^{ic\chi/2 }  \\
\sqrt{R}e^{-ic\chi/2 } & -\sqrt{D}
\end{pmatrix}.
\label{A2}
\end{equation}
In order to evaluate the current we now need to substitute the wave functions $\Psi^+_{k,c;L,\mu}$ and $\Psi_{k,c;L,\mu}(t)$ into Eq. \eqref{T8}.
At times $t$ sufficiently far from the time moments $t_{0,k,c}$ and $t_{2\pi,k,c}$ we may employ the basis of adiabatic wave functions 
\begin{equation}
\Psi_{k,c;L,\mu} (t)
=
\sum_{\nu}
\mathcal{S}_{k,c;\nu,\mu}(t)
\sqrt{|\sin[\chi(t)/2]|}
\psi_{k, c; L,\mu}(t),
\label{psipsi}
\end{equation}
where $\psi_{k, c; L,\mu}(t)$ is defined in Eq. \eqref{adiabatic}. The elements of the scattering matrix $\mathcal{S}_{k,c;\nu,\mu}(t)$ \eqref{Smatrix} $r_{k,c}(t)$ and $d_{k,c}(t)$ describe the process of Landau-Zener tunneling between Andreev levels
in the vicinity of the anticrossing point $\chi=4\pi k + \pi c$. It is important to emphasize that each $2\pi$ phase interval can be treated separately from the other ones, since the system evolution within one such interval does not depend on that for all the remaining intervals.

Substituting Eq. \eqref{psipsi} into Eq. \eqref{T8}, we immediately recover the result \cite{KZ26} for the Josephson current  generalized to arbitrary temperatures, i.e.
\begin{multline}
I_J[\chi(t)]=
-2e\tanh\dfrac{|\Delta|}{2T}\dfrac{\partial \varepsilon_A(\chi)}{\partial \chi}
\Biggl[
[1 - 2|d(t)|^2]
\\+
2  \sqrt{R} |r(t)| |d(t)|  \tan [\chi(t)/2]
\\\times
\cos\left( \delta(t)  + 2 \int_{t_{\pi,k,c}}^t\varepsilon_A[\chi(\tilde t)] d \tilde t \right)
\Biggr],
\label{T+}
\end{multline}
with the phase $\delta(t)$ is defined as
\begin{equation}
\delta(t) = \pi - \arg d(t) - \arg r(t).
\label{deltat}
\end{equation}
Here and below we suppress the interval indices for the scattering amplitudes replacing $r_{k,c}(t)$ and $d_{k,c}(t)$ respectively by $r(t)$ and $d(t)$. The corresponding values $k$ and $c$ are uniquely determined by the condition $t_{0,k,c}<t<t_{2\pi,k,c}$.

Within a given $2\pi$ interval $0< \tilde \chi <2\pi$ with
$\tilde\chi=\chi-2\pi n$ being the "reduced" phase, and sufficiently far from the anticrossing region $\tilde \chi \approx \pi$ we have $r(t)= 0$ and $|d(t)|= 1$ for $0<\tilde \chi <\pi$ and \cite{Z32,Vit96,Iv23}
\begin{gather}
d = - e^{-\pi s},
\label{d}
\\
\delta=
 -\dfrac{\pi}{4} + s\ln \frac{s}{e}  -  \arg \Gamma(is)
\label{delta}
\end{gather}
for $\pi< \tilde \chi <2\pi$. Here $\Gamma(x)$ is the Euler gamma function and
\begin{equation}
s=\frac{R|\Delta|}{2eV(t_\pi)}
\end{equation}
is the Landau-Zener adiabaticity parameter evaluated at the time $t_\pi$ defined by the condition $\tilde\chi(t_\pi)=\pi$. For a time-dependent voltage bias, different avoided crossings generally correspond to different values of $V(t_\pi)$ and are therefore characterized by different values of $s$, $d$, and $\delta$.

The above result fully determines the non-dissipative contribution to the total current $I(t)$ as a function of the Josephson phase $\chi (t)$ at non-zero external voltage bias. As we already demonstrated before \cite{KZ26},  Eq. \eqref{T+} accounts for non-trivial consequences of Landau-Zener tunneling, such as coherent oscillations on CPR within a given $2\pi$ phase interval which occur after passing the anticrossing point $\chi = \pi + 2\pi n$ and may become pronounced for $R \ll 1$ and small $eV \ll |\Delta |$.

So far we assumed that the phase $\chi(t)$ remains sufficiently far from the points $\chi=2\pi n$. The situation changes as soon as the phase variable approaches one of the interval ends. In this case the effective Hamiltonian description breaks down and the wave functions $\Psi_{k,c;X,\mu}(t)$ should be replaced by the corresponding Airy-function solutions, see Appendices \ref{appA} and \ref{appB}. As a result, the wave functions are not anymore localized within a given $2\pi$ interval and acquire tails extending into the neighboring interval. Hence, the current $I[\chi (t)]$ with $\chi (t)$ in the vicinity of any interval boundary should receive extra contributions associated with the two adjacent intervals.

As we already discussed, these extra contributions to the current turn out to be dissipative as they arise from quasiparticle transitions between subgap Andreev states and the continuum. In order to find the dissipative current $I_{\rm diss} [\chi (t)]$ it is necessary to evaluate the kernel of the inverse operator $(W + a)^{-1}$ provided the time arguments $t$ and $t'$ are allowed to approach the ends of a given $2\pi$ phase interval. As due to the causality condition we always have $t' < t$, three different situations may be realized: (i) $t'$ is close to $t_{0,k,c}$ while $t$ is close to $t_{2\pi,k,c}$, (ii) both $t$ and $t'$ are close to $t_{0,k,c}$ and (iii) both $t$ and $t'$ are close to $t_{2\pi,k,c}$. 

In the case (i) the wave function $\Psi_{k,c;L,\mu}(t)$ can still be expressed in terms of Eq. \eqref{psipsi} with the adiabatic wave functions $\psi_{k, c; L,\mu}(t)$ replaced by their Airy-function counterparts, see Eqs.~\eqref{tpsip} and \eqref{tpsim}. Substituting the resulting expression for $\Psi_{k,c;L,\mu}(t)$ into Eq. \eqref{T8}, one finds
\begin{multline}
I_1[\chi(t)]=
-
e D |\Delta|
\left(
\dfrac{\dot \chi}{2D|\Delta|}
\right)^{1/3}
[h_{+} - h_{-}]
\\\times
[1 - 2|d_{k,c} (t)|^2]
|f_{L,+}(\tau)|^2,
\label{I3}
\end{multline}
where the function $f_{L,+}(\tau)$ is defined in Eq. \eqref{fLAiry} and the variable $\tau$ represents the dimensionless "distance" to the time moment $t_{2\pi,k,c}$
\begin{equation}
t- t_{2\pi,k,c} = 4 \tau /(2D |\Delta| \dot \chi^2 )^{1/3}.
\end{equation}

Physically, the contribution \eqref{I3}  describes the process of a quasiparticle transfer from the continuum to the discrete Andreev state at $t'\sim t_{0,k,c}$ with subsequent evolution of this state according to the Schr\"odinger-like equation \eqref{shrodLR} followed by a final release of the quasiparticle back into the continuum at $t\sim t_{2\pi,k,c}$.

The two remaining contributions (ii) and (iii) describe local transitions between the Andreev state and the continuum near the gap edge. Since in both cases the propagation from $t'$ to $t$ is always confined to the vicinity of the point where the Andreev level merges with the continuum, the system never reaches the avoided crossing region $\tilde \chi \approx \pi$. Hence, the sum of the contributions (ii) and (iii) -- denoted below as $I_{23}$ -- is insensitive to Landau-Zener tunneling being independent of the tunneling probability $|d|^2$.

This observation allows one to reconstruct $I_{23}$ without evaluating it explicitly by means of the following trick.  Notice that in the limit of vanishing Landau-Zener tunneling, $d\to 0$, the total current reduces to the standard adiabatic form
\begin{equation}
I_{\mathrm{ad}} \equiv I_J (d\to 0)=
-2e\tanh\dfrac{|\Delta|}{2T}\dfrac{\partial \varepsilon_A(\chi)}{\partial \chi}.
\label{Iad}
\end{equation}
The same result \eqref{Iad} should be reproduced by evaluating the total current $I_1+I_{23}$ at $d \to 0$. Since $I_{23}$ is independent of $|d|^2$ and $I_1$ is established by Eq.~\eqref{I3}, one can formally set $|d_{k,c}(t)|^2=0$ in this equation and subtract the resulting expression for $I_1$ from Eq.  \eqref{Iad}. As a result, we arrive at the expression for $I_{23}$ which identically coincides with that obtained by means of an explicit calculation leading to Eq.~\eqref{I23} (see Appendix \ref{appB} for details).

Hence, we recover the total current $I[\chi(t)]$ in the form of Eq. \eqref{123}. What remains is to subtract the current $I_J[\chi (t)]$ \eqref{T+} from the total current $I[\chi(t)]$ \eqref{123} which -- according to Eq. \eqref{sum} -- immediately yields the dissipative contribution to the current
\begin{multline}
I_{\rm diss}[\chi(t)]=
4e D |\Delta| \tanh\dfrac{|\Delta|}{2T}
\left(\dfrac{eV}{D|\Delta|}\right)^{1/3}
\\\times
|d|^2
g\left[\left(\dfrac{D|\Delta|}{eV}\right)^{1/3}\dfrac{|\delta \chi(t)|}{4}\right],
\label{Idiss3}
\end{multline}
where  $|\delta\chi(t)|$ denotes the distance between $\chi(t)$ and the nearest point $2\pi n$, $g(\tau)$ is the universal function defined as
\begin{multline}
\label{gdef}
g(\tau)
= |f_{L,+}(\tau)|^2 - |\tau| \theta(-\tau)
\\=
\Biggl[
\pi\tau^2 \left|\Ai \left(\tau^2 e^{i\pi /3}\right)\right|^2
+
\pi\left|\Ai' \left(\tau^2 e^{i\pi /3}\right)\right|^2
-\dfrac{|\tau|}{2}
\biggr]
\end{multline}
and
\begin{equation}
|d|^2
=
\exp\left(
-\dfrac{\pi R |\Delta|}{eV}
\right)
\label{Idissf}
\end{equation}
is the Landau-Zener tunneling probability at the most recent avoided crossing traversed by the system. For brevity we suppress the indices on $d$.
We observe that the junction transparency $D$ enters the dimensionless parameter $(eV/D|\Delta|)^{1/3}$ in Eq. \eqref{Idiss3}. This is
due to the behavior of the Andreev level near the gap edge $\varepsilon_{A}(\chi)\approx |\Delta|\left[1 - D\chi^2/8\right]$ for $|\chi|\ll 1$.

Note that the dissipative contribution to the current \eqref{Idiss3} -- similarly to a non-dissipative one in Eq. \eqref{T+} -- contains the probability of Landau-Zener tunneling $|d (t)|^2$. Its origin is transparent: For a junction with non-zero $R$, a quasiparticle "climbing" in energy from $-|\Delta|$ to $|\Delta|$ inevitably performs Landau-Zener tunneling between the two Andreev levels.

For $g(\tau)$ we obtain the asymptotic behavior
\begin{equation}
g(\tau)
=
\begin{cases}
g(0) - \dfrac{|\tau|}{2} + \dfrac{\tau^2}{12g(0)} + \mathcal{O}(\tau^4), & |\tau| \ll 1,
\\
\dfrac{1}{64 |\tau|^5}-
\dfrac{105}{2^{12}|\tau|^{11}}
+
\mathcal{O}(1/|\tau|^{17}), & |\tau| \gg 1,
\end{cases}
\end{equation}
where $g(0)=\Gamma(5/6)/(2^{4/3} 3^{1/6}\sqrt{\pi}) \approx 0.210$.
An accurate interpolation for $g(\tau)$ is provided by the function
\begin{equation}
\tilde g(\tau) = \dfrac{g(0)}{ 1 + \dfrac{|\tau|}{2g(0)} + \dfrac{\alpha|\tau|^2}{6g^2(0)}+64 g(0)|\tau|^5}.
\label{tildeg}
\end{equation}
For $\alpha = 1$, the function $\tilde g(\tau)$ exactly reproduces the first three terms of the Taylor expansion of $g(\tau)$ as well as its leading asymptotic behavior at $|\tau| \to \infty$. Even better agreement at intermediate $\tau$ is achieved for $\alpha = 1.465$. In this case the absolute and relative errors over the entire range of $\tau$ do not exceed $0.0025$ and $0.05$, respectively. 

\begin{figure}
\begin{center}
\includegraphics[width=80mm]{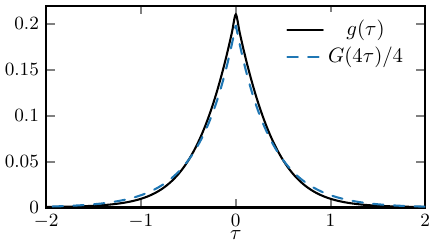}
\end{center}
\caption{Universal functions $g(\tau)$ and $G(\tau)$ (with $b = 2.5$) defined respectively in Eq.~\eqref{gdef} and \eqref{Gt}.}
\label{Idiss-fig}
\end{figure}

The universal function $g(\tau)$ \eqref{gdef} is displayed in Fig.~\ref{Idiss-fig}. The function  $\tilde g(\tau)$ \eqref{tildeg} with $\alpha = 1.465$ is not shown since it is indistinguishable from $g(\tau)$ within the line thickness. We observe that the function $g(\tau)$ is peaked at around $\tau = 0$ and decays rather quickly with increasing $|\tau |$ being about an order of magnitude smaller than $g(0)$ already at $|\tau | \approx 1$. In other words, the dissipative current \eqref{Idissf} remains appreciable for the phase values 
\begin{equation}
|\delta \chi(t)| \ll 4\left(\dfrac{eV}{D|\Delta|}\right)^{1/3}
\label{chidiss}
\end{equation}
and practically vanishes outside this interval.

Note that previously the current $I_{\rm diss} [\chi (t)]$ was evaluated numerically \cite{GZ3} in the limit of fully transparent junctions with $D \to 1$ and at $T \ll |\Delta|$. The dependencies calculated at different voltage values essentially collapsed onto a single curve which was approximated by the formula \cite{GZ3}
\begin{equation}
I_{\rm diss}[\chi(t)]=e|\Delta|
\left(\dfrac{eV}{|\Delta|}\right)^{1/3}
G\left[\left(\dfrac{|\Delta|}{eV}\right)^{1/3} |\delta \chi(t)|
\right], 
\label{diss}
\end{equation}
where, as before, $\delta \chi(t)$ denotes the distance between $\chi(t)$ and the nearest point $2\pi n$ and the universal function $G(\tau )$ was chosen in the form
\begin{equation}
G(x)=\frac{b}{\pi}\exp (-0.27b |x|)
\label{Gt}
\end{equation}
with the parameter equal to $b \simeq 2.5$  for $eV/|\Delta| \lesssim 0.01$ and to $b \simeq 2.7 $  for $eV/|\Delta| \gtrsim 0.1$. The function
$G(\tau)$ is also plotted in Fig.~\ref{Idiss-fig} for comparison.

\begin{figure}
\begin{center}
\includegraphics[width=8cm]{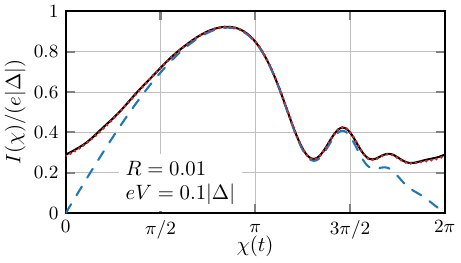}
\end{center}
\caption{The current $I$ in a single-channel junction as a function of $\chi(t)$ for $R=0.01$, $eV=0.1|\Delta|$ and $T \to 0$.
The black solid line shows a numerically exact result. The blue dashed line corresponds to the Josephson current $I_J$ \eqref{T+}, with $r(t)$ and $d(t)$ obtained from the solution for the Landau-Zener problem \cite{KZ26}.
The red dotted line indicates the total current \eqref{sum} which includes $I_J$ and a dissipative correction $I_{\rm diss}$ defined in Eq.~\eqref{Idiss3}.}
\label{Ichi-R001-V010-paper2-fig}
\end{figure}

\begin{figure}
\begin{center}
\includegraphics[width=8cm]{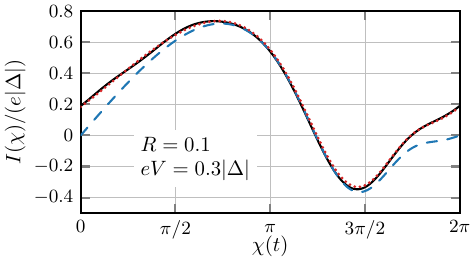}
\end{center}
\caption{The same as in Fig.~\ref{Ichi-R001-V010-paper2-fig} for $R=0.1$ and $eV = 0.3 |\Delta|$.}
\label{Ichi-R01-V03-paper-fig}
\end{figure}

\begin{figure}
\begin{center}
\includegraphics[width=8cm]{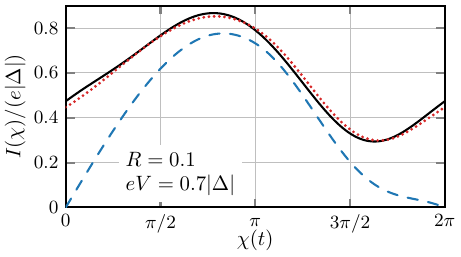}
\end{center}
\caption{The same as in Fig.~\ref{Ichi-R001-V010-paper2-fig} for $R=0.1$ and $eV = 0.7 |\Delta|$.}
\label{Ichi-R01-V07-paper-fig}
\end{figure}

\begin{figure}
\begin{center}
\includegraphics[width=8cm]{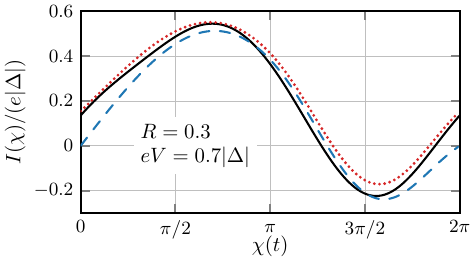}
\end{center}
\caption{The same as in Fig.~\ref{Ichi-R001-V010-paper2-fig} for $R=0.3$ and $eV = 0.7 |\Delta|$.}
\label{Ichi-R03-V07-paper-fig}
\end{figure}

In Figs.~\ref{Ichi-R001-V010-paper2-fig}-\ref{Ichi-R03-V07-paper-fig} we display CPR for a superconducting junction out of equilibrium derived above at different reflection coefficients $R$ and applied bias voltages $V$. Our analytical results are also supplemented by numerical ones
obtained with the aid of exact recurrence relations \cite{KZ} equivalent to the expression \eqref{T3} in the limit of a constant in time bias voltage. The numerical code employed in these calculations is publicly available at \cite{Code}.

All essential features of CPR predicted above can be observed in Fig. ~\ref{Ichi-R001-V010-paper2-fig}. For instance, traces of coherent oscillations at $\pi <\chi <2\pi$ remain clearly visible (although these oscillations can be much more pronounced for even smaller voltage values, cf. \cite{KZ26}). Away from the points $\chi = 2\pi n$, the current is well described by the Josephson term $I_J(\chi)$ ~\eqref{T+}. At the same time,  as the phase gets closer to the points $\chi = 2\pi n$ pronounced deviations between $I_J(\chi)$ and numerically exact results appear. However, this discrepancy disappears completely provided one also takes into account the dissipative contribution $I_{\rm diss}(\chi)$ ~\eqref{Idiss3} originating from quasiparticle transitions into the continuum. The total current $I(\chi)=I_J(\chi)+I_{\rm diss}(\chi)$  turns out to be in excellent agreement with numerical results. The same situation is also observed in Fig. \ref{Ichi-R01-V03-paper-fig}.

Remarkably, our theory appears to work well even beyond its strict applicability range $eV \ll |\Delta |$. For instance, in Fig. \ref{Ichi-R01-V07-paper-fig} we display CPR for $eV=0.7|\Delta|$ and $R=0.1$. While the dependence $I_J(\chi)$ ~\eqref{T+} considerably deviates from the numerical curve, a good agreement between our analytical and numerical results is immediately restored by adding the dissipative contribution $I_{\rm diss}(\chi)$ ~\eqref{Idiss3}. Moreover, a reasonably good agreement is observed even at bigger values of $R=0.3$, see Fig. \ref{Ichi-R03-V07-paper-fig}.

To complete our analysis we evaluate the average total current $\overline{I}$ as a function of the applied voltage $V$. Assuming for simplicity that the probability $|d|^2$ equals to zero before the Landau-Zener tunneling event (i.e. at $t < t_{\pi,k,c}$) and to the time-independent value  \eqref{Idissf} after this event, after proper time averaging of the total current \eqref{sum} we arrive at the sum of the two terms
\begin{equation}
\overline{I} =
\dfrac{2}{\pi} e|\Delta| |d|^2
\Biggl[ 1
+
\dfrac{\Gamma(2/3)}{2^{2/3} 3^{1/3}}
D \left(\dfrac{eV}{D|\Delta|}\right)^{2/3}
\Biggr]
\tanh\dfrac{|\Delta|}{2T},
\label{IV}
\end{equation}
where $\Gamma(2/3)/(2^{2/3} 3^{1/3}) \approx 0.591$. The first term in this sum originates from the Josephson contribution to the current $I_J[\chi (t)]$ and reduces to the result \cite{Uwe} for fully transparent junctions with $R \to 0$. The last term in Eq. \eqref{IV} is obtained by averaging the dissipative current $I_{\rm diss}[\chi (t)]$ over time.  It matches exactly with the analogous contribution derived earlier \cite{GZ1,GZ2} in the limit $D \to 1$.

\section{Discussion}
\label{secdiscussion}

Recent developments demonstrate that -- despite decades of investigations -- until now ac Josephson effect was not yet fully understood beyond the tunneling limit. In particular, at low voltages $eV \ll |\Delta |$ and for highly transparent junctions with $R \ll 1$ several novel features have been discovered. Firstly, coherent oscillations on CPR of such junctions have been predicted \cite{KZ26} as a result of Landau-Zener tunneling between the two subgap Andreev bound states. Secondly, as we demonstrated here, a rigorous treatment of quasiparticle dynamics near the superconducting gap edge yields an additional sub-Ohmic phase-dependent dissipative contribution to CPR which remains appreciable even deep in the subgap regime. In this regime, the above new features result in a rather non-trivial CPR defined by Eqs. \eqref{sum}, \eqref{T+} and \eqref{Idiss3} and illustrated in Figs. ~\ref{Ichi-R001-V010-paper2-fig}-\ref{Ichi-R03-V07-paper-fig}.

\begin{figure}
\begin{center}
\includegraphics[width=8cm]{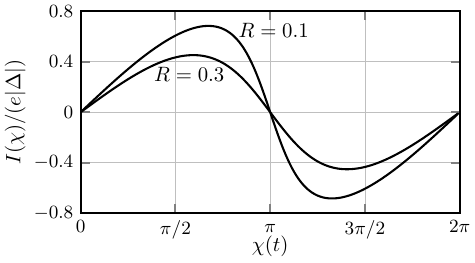}
\end{center}
\caption{CPR at $T \to 0$ plotted for $R=0.3$, $eV = 0.1 |\Delta|$ and for $R=0.1$, $eV = 0.03 |\Delta|$. Both curves indicate numerically exact results. The dependence in Eq. \eqref{Iad}  coincides with these curves within the line thickness.}
\label{Ichi-eq-paper-fig}
\end{figure}

At sufficiently low voltages,
\begin{equation}
eV \ll  \pi R |\Delta|,
\label{ineq}
\end{equation}
the probability of Landau-Zener tunneling $|d|^2$ \eqref{Idissf} remains exponentially small and can be neglected. In this case
both oscillating and dissipative contributions to the current are essentially suppressed and CPR reduces to a simple form \eqref{Iad}.
This situation is illustrated in Fig.~\ref{Ichi-eq-paper-fig} demonstrating an excellent agreement between the dependence in Eq.~\eqref{Iad} and numerically exact results.

Note that beyond the tunneling limit and at $T >0$ Eq.~\eqref{Iad} does not reduce to the equilibrium CPR of Josephson junctions \cite{KO,GKI} even at $V \to 0$. This is because for our problem the continuous spectra effectively act as energy reservoirs for quasiparticles occupying both Andreev levels and, hence, the filling factors for these levels are controlled by the Fermi function at energies $\varepsilon = \pm |\Delta |$ rather than at $\varepsilon = \pm \varepsilon_A(\chi )$ (as in equilibrium). In order to cross over to the equilibrium Josephson current at small voltages it is necessary to introduce some inelastic relaxation mechanism that would be "faster" than the sweeping rate of the phase $\chi (t)$ at a given $V$. Such a mechanism is not considered within our present analysis.

The simple equation \eqref{Iad} stops working as soon as the inequality \eqref{ineq} is violated. Then Landau-Zener tunneling comes into play
giving rise to both coherent oscillations in CPR described by Eq. \eqref{T+} and the phase-dependent dissipative current in Eq. \eqref{Idiss3}.
Note that in order to properly account for all these effects the standard quantum mechanical analysis would not be sufficient for a number of reasons. Conceptually, Andreev bound states cannot be treated as ordinary single particle levels in a potential well since such states correspond to a superposition of quasiparticles and holes defined within the frames of a non-trivial many-body problem. On top of that, quantum dynamics of Andreev levels becomes non-unitary and dissipative as soon as a non-zero bias voltage is applied to the junction. Hence, it is necessary to develop a fully microscopic quantum description of non-equilibrium evolution of Andreev states that would include dissipative effects. This task is accomplished in our present work.

What remains is to comment on the relation between our present approach and the one based on the standard physical picture of MAR \cite{MAR} which does not involve Andreev levels. Although these two approaches may seem conceptually different, both of them technically follow from the same general expression for the current \eqref{T3} and both of them describe the same physical processes from complementary perspectives.

In order to see this let us recall that after one full MAR cycle (two traversals across the junction accompanied by two Andreev reflections)
the charge $2e$ is transferred between two superconducting electrodes and a quasiparticle gains the energy $\delta E=2eV.$ Within the Andreev level picture, the same charge $2e$ is transferred during the time interval $\delta t =1/\partial_{\chi}E(\chi)$ since a singly occupied
Andreev level carries the current $I=2e\partial_{\chi}E(\chi)$. Evaluating the quasiparticle energy gain during the time $\delta t$ we again obtain
$$
\delta E = E[\chi (t+\delta t)]-E[\chi (t)]\simeq \partial_{\chi}E(\chi)\dot \chi \delta t = 2eV.
$$
Thus, our present approach describes the system evolution in the time domain whereas the MAR-based picture deals with the same evolution in energy space.

The choice of either approach is then just a matter of convenience. For instance, the subharmonic gap structure on the $I-V$ curve \cite{Chalmers,Madrid} naturally follows from the approach based on the physical picture of MAR. On the other hand, this picture appears less transparent for the effects studied here, in particular if one wants to go beyond the perturbative treatment in $R$ \cite{GZ4}. We anticipate that 
a proper combination of both approaches might be appropriate in more complicated systems \cite{KZ} which are beyond the frames of the present work.

\appendix

\section{}
\label{appA}

Let us assume that the time variable $t$ belongs to the interval $t_{0, k, c} < t < t_{2\pi, k, c}$ and remains sufficiently far from its boundaries. Then in order to evaluate the Josephson current one needs to reconstruct the wave function $\Psi^+_{k, c; R,\mu}(t')$ at $t'$ close to the time moment $t_{0, k, c}$. For this purpose we will first seek for a general solution of the integral equation (cf. Eq. \eqref{PsiRW})
\begin{equation}
F_{R}(t') W_{c}[\chi(t')] +  \int_{t'}^{\infty} dt_1 F_{R}(t_1) a(t_1-t') = 0,
\label{FR}
\end{equation}
for $t' \sim t_{0, k, c}$, and then construct a solution that matches $\Psi^+_{k, c; R,\mu}(t')$ away from this point.

Close to $t_{0,k,c}$, the matrix $W_{c}[\chi(t')]$ can be approximately expressed in the form
\begin{equation}
W_{c}[\chi(t')]
\approx
\begin{pmatrix}
c\sqrt{D} & \sqrt{R} \\
\sqrt{R} & -c\sqrt{D}
\end{pmatrix}
-
\dfrac{i}{2}\sqrt{D}\dot \chi (t' - t_{0, k, c}).
\label{W-ct0}
\end{equation}
Substituting this expansion into Eq.~\eqref{FR}, we obtain the following homogeneous integral equation for $F_R$
\begin{equation}
F_R
\left[
\begin{pmatrix}
c\sqrt{D} & \sqrt{R} \\
\sqrt{R} & -c\sqrt{D}
\end{pmatrix}
-
\dfrac{i}{2}\sqrt{D}\dot \chi (t' - t_{0, k, c})
+
a
\right] =0.
\label{FR2}
\end{equation}
We can represent the solution of Eq.~\eqref{FR2} as
\begin{equation}
F_R =
\sum_{\mu=\pm}
F_{R,\mu}
\begin{pmatrix}
\sqrt{R}  \\ \mu - c \sqrt{D}
\end{pmatrix}
\mu \dfrac{\mu  +  c \sqrt{D}}{2R}
\begin{pmatrix}
\sqrt{R}  & \mu - c \sqrt{D}
\end{pmatrix}.
\label{A6}
\end{equation}
where function $F_{R,\mu}$ obeys the integral equation
\begin{equation}
F_{R,\mu}
\left[
\mu
-
\dfrac{i}{2}\sqrt{D}\dot \chi (t' - t_{0, k, c})
+
a
\right] =0.
\end{equation}
In order to analyze this equation at times $t' \sim t_{0,k,c}$ it would be convenient to factor out the rapidly oscillating phase and to write
\begin{equation}
F_{R,\mu}(t') = e^{i\mu|\Delta|(t' - t_{0, k, c})} f_{R,\mu}(\tau),
\end{equation}
where the dimensionless time $\tau$ is defined by the equation
\begin{equation}
t' - t_{0, k, c} = 4 \tau /(2D |\Delta| \dot \chi^2 )^{1/3}.
\end{equation}
Expressing $f_{R,\mu}(\tau)$ in terms of the Fourier integral
\begin{equation}
f_{R,\mu}(\tau) = \int \dfrac{d \omega}{2\pi}
f_{R,\mu}(\omega) e^{i\mu \omega \tau}
\label{fmuf+}
\end{equation}
and substituting Eq.~\eqref{fmuf+} into the equation for $F_{R,\mu}$, we obtain
\begin{multline}
\Biggl[\mu +
\mu \dfrac{2\sqrt{D}\dot \chi}{(2D |\Delta| \dot \chi^2 )^{1/3}}
\dfrac{d}{d\omega}
\\+
a[\mu|\Delta| + \mu\omega(2D |\Delta| \dot \chi^2 )^{1/3}/4]\Biggr]
f_{R,\mu}(\omega)=0.
\end{multline}

In order to proceed, let us expand $a(\varepsilon)$ in the vicinity of $\varepsilon=\mu|\Delta|$ as
\begin{equation}
a(\mu|\Delta| + \mu \delta \varepsilon) \approx -\mu +
\mu
\dfrac{\sqrt{2\delta\varepsilon}}{\sqrt{|\Delta|}},
\label{aexp}
\end{equation}
where the branch of $\sqrt{\delta\varepsilon}$ is chosen analytic in the upper (lower) half-plane for $\mu=+1$ ($\mu=-1$). Within this approximation, the equation for $f_{R,\mu}(\omega)$ reduces to
\begin{equation}
\left[
\dfrac{d}{d\omega}
+
\dfrac{1}{2}
\sqrt{\omega}\right]
f_{R,\mu}(\omega)=0,
\end{equation}
which yields
\begin{equation}
f_{R,\mu}(\omega) =
\dfrac{\sqrt{\pi}}{2}
\exp\left(
-
\dfrac{1}{3}
\omega^{3/2}
\right).
\label{fmusol+}
\end{equation}
The overall normalization in Eq. \eqref{fmusol+} is chosen for later convenience. One should note that although the right-hand side of Eq. \eqref{fmusol+} does not explicitly depend on $\mu$, this dependence nevertheless enters indirectly through the choice of the branch of $\omega^{3/2}$, which is set to be analytic in the upper (lower) half-plane for $\mu=+1$ ($\mu=-1$).

Note that the function equivalent to that in Eq. \eqref{fmuf+} with $f_{R,\mu}(\omega)$ defined in Eq. \eqref{fmusol+} also appears in the analysis \cite{Houset} of quasiparticle transitions between Majorana bound states and the continuum in fully transparent ($D=1$) topological Josephson junctions. On the other hand, for any $D<1$ Majorana bound states do not touch the continuum and the situation \cite{Houset} becomes entirely different from that considered here.

Comparing $f_{R,\mu}(\omega)$ with the analogous solution near $t_{2\pi,k,c}$, we get
\begin{equation}
f_{R,\mu}(\tau) = f_{L,\mu}(-\tau),
\end{equation}
where the function $f_{L,\mu}$ can be expressed in terms of Airy function, see Eq.  \eqref{fLAiry} below.

By definition, at $t' \sim t_{0,k,c}$ the wave function
$\Psi^+_{k,c;R,\mu}(t')$ should coincide with the combination
$\sqrt{|\sin[\chi(t')/2]|}\,\psi^+_{k,c;R,\mu}(t')$, where the adiabatic wave functions read
\begin{widetext}
\begin{gather}
\psi^+_{k, c;R, +}(t')
=
\dfrac{
\begin{pmatrix}
\sqrt{R} \\ 1 - c\sqrt{D}
\end{pmatrix}^T
}{
\sqrt{2}\sqrt{1 - c\sqrt{D}}}
\exp\left(
i \int_{t_{\pi, k, c}}^{t_{0, k, c}}\varepsilon_A[\chi(\tilde t)] d \tilde t
\right)
\exp\left(
i|\Delta|(t'-t_{0, k, c}) - i \dfrac{4 \tau^3}{3}
\right),
\\
\psi^+_{k, c;R, -}(t')
=
-i
\dfrac{
\begin{pmatrix}
1 - c\sqrt{D} \\ -\sqrt{R}
\end{pmatrix}^T
}{
\sqrt{2}\sqrt{1 - c\sqrt{D}}}
\exp\left(
-i \int_{t_{\pi, k, c}}^{t_{0, k, c}}\varepsilon_A[\chi(\tilde t)] d \tilde t
\right)
\exp\left(
-i|\Delta|(t'-t_{0, k, c}) + i \dfrac{4 \tau^3}{3}
\right),
\end{gather}

Comparing these expressions with the asymptotics of $f_{L,\mu}(\tau)$,
we obtain the matching conditions
\begin{multline}
\sqrt{|\sin[\chi(t')/2]|} \psi^+_{k, c; R, +}(t')
\Leftrightarrow
\dfrac{
\begin{pmatrix}
\sqrt{R} & 1 - c\sqrt{D}
\end{pmatrix}
}{
\sqrt{2}\sqrt{1 - c\sqrt{D}}}
\sqrt{2}
\left(
\dfrac{\dot \chi}{2D|\Delta|}
\right)^{1/6}
e^{-i\pi/4}
f_{L,+}(-\tau)
\times\\\times
\exp\left(
-i \int_{t_{\pi, k, c}}^{t_{0, k, c}}\varepsilon_A[\chi(\tilde t)] d \tilde t
\right)
\exp\left[
-i|\Delta|(t'-t_{0, k, c})\right],
\label{tpsip+}
\end{multline}
and
\begin{multline}
\sqrt{|\sin[\chi(t')/2]|} \psi^+_{k, c;R,-}(t')
\Leftrightarrow
-i
\dfrac{
\begin{pmatrix}
1 - c\sqrt{D} & -\sqrt{R}
\end{pmatrix}
}{
\sqrt{2}\sqrt{1 - c\sqrt{D}}}
\sqrt{2}
\left(
\dfrac{\dot \chi}{2D|\Delta|}
\right)^{1/6}
e^{i\pi/4}
f_{L,-}(-\tau)
\times\\\times
\exp\left(
i \int_{t_{\pi, k, c}}^{t_{0, k, c}}\varepsilon_A[\chi(\tilde t)] d \tilde t
\right)
\exp\left[
i|\Delta|(t'-t_{0, k, c})\right].
\label{tpsim+}
\end{multline}
Note that both sides of these matching conditions have identical asymptotic behavior in the overlap region
\begin{equation}
t' > t_{0, k, c},
\quad
(2D |\Delta| \dot \chi^2 )^{1/3} |t' - t_{0, k, c}| \gg 1,
\quad
|\chi(t') - \chi(t_{0, k, c})| \ll 1.
\end{equation}

\end{widetext}

\section{}
\label{appB}

In Appendix \ref{appA} we performed our analysis assuming that the phase $\chi(t)$ remains not too close to the points $\chi = 2\pi n$ in which case the current across our junction is mainly determined by the dissipativeless Josephson contribution. Below we relax this assumption and allow the phase $\chi(t)$ to approach the values $2\pi n$ which is necessary to reconstruct the dissipative contribution to the current.

Our goal is to determine the inverse operator $(W+a)^{-1}$ beyond the validity domain of Eq. \eqref{invW}. Within the approximation employing the Schr\"odinger-like equation \eqref{shrodLR} describing both adiabatic evolution and Landau-Zener tunneling, Eq. \eqref{invW} represents $(W+a)^{-1}$ as a sum of contributions labeled by the indices $(k,c)$, each of which is strictly confined to the corresponding time interval $t_{0,k,c}<t<t_{2\pi,k,c}$. As discussed above, this approximation breaks down in the vicinity of the boundaries of such time intervals. In this Appendix we perform a more accurate analysis and demonstrate that each $(k,c)$ contribution should be viewed as a result of extending the corresponding term in Eq. \eqref{invW} beyond its formal validity domain. Rather than being terminated abruptly at the interval boundaries, these terms decay rapidly outside the corresponding time intervals.

It follows immediately that, provided the observation time $t$ approaches the interval boundary, the contributions from both adjacent intervals must be taken into account as well. For definiteness, let $t$ approach $t_{2\pi,k,c}$. In this case three different contributions to the current occur. The first one originates from $(k,c)$-term in Eq. \eqref{invW} with $t'$ remaining close to $t_{0,k,c}$. The second one is obtained from extending the same $(k,c)$-term to the region where both $t$ and $t'$ are close to $t_{2\pi,k,c}$. The third one follows from the extending the neighboring term with indices $(k+1/2+c/2,-c)$ to the region where both $t$ and $t'$ are close to $t_{0,k+1/2+c/2,-c}$. Since $t_{0,k+1/2+c/2,-c}=t_{2\pi,k,c}$, the latter two terms contribute in the same time domain. These three contributions -- also discussed in Sec. \ref{secccurrent} -- are explicitly evaluated below.

\subsection{The time variables $t$ and $t'$ are close to respectively  $t_{2\pi,k,c}$ and $t_{0,k,c}$}

Since the behavior of the wave function $\Psi_{k,c;R,\mu}(t')$ in the vicinity of $t_{0,k,c}$ was already established in Appendix \ref{appA}, it only remains to determine $\Psi_{k,c;L,\mu}(t)$ at times $t$ close to $t_{2\pi,k,c}$. For this purpose it suffices to resolve the following homogeneous integral equation
\begin{equation}
[W_{c} +  a] F_L =0.
\label{FL}
\end{equation}

Near $t_{2\pi,k,c}$ one can approximately write
\begin{equation}
W_{c}[\chi(t)] \approx
\begin{pmatrix}
-c\sqrt{D} & \sqrt{R} \\
\sqrt{R} & c\sqrt{D}
\end{pmatrix}
+
\dfrac{i}{2}\sqrt{D}\dot \chi (t - t_{2\pi, k, c}).
\label{B2}
\end{equation}
Combining Eqs .~\eqref{FL} and \eqref{B2}, we obtain 
\begin{equation}
\left[
\begin{pmatrix}
-c\sqrt{D} & \sqrt{R} \\
\sqrt{R} & c\sqrt{D}
\end{pmatrix}
+
\dfrac{i}{2}\sqrt{D}\dot \chi (t - t_{2\pi, k, c})
+
a
\right] F_L =0.
\label{WF2}
\end{equation}
We can represent the solution of Eq.~\eqref{WF2} in the form
\begin{equation}
F_L =
\sum_{\mu=\pm}
\begin{pmatrix}
\sqrt{R}  \\ \mu + c \sqrt{D}
\end{pmatrix}
F_{L,\mu} \mu \dfrac{\mu  -  c \sqrt{D}}{2R}
\begin{pmatrix}
\sqrt{R}  & \mu + c \sqrt{D}
\end{pmatrix},
\end{equation}
where the coefficients $F_{L,\mu}(t)$ satisfy the scalar integral equation
\begin{equation}
\left[
\mu
+
\dfrac{i}{2}\sqrt{D}\dot \chi (t - t_{2\pi, k, c})
+
a
\right] F_{L,\mu} =0.
\end{equation}
As before, in order to analyze this equation in the vicinity of $t_{2\pi,k,c}$ it is convenient to extract the rapidly oscillating phase as
\begin{equation}
F_{L,\mu}(t) = e^{-i\mu|\Delta|(t - t_{2\pi, k, c})} f_{L,\mu}(\tau),
\end{equation}
where $\tau$ is defined from
\begin{equation}
t - t_{2\pi, k, c} = 4 \tau /(2D |\Delta| \dot \chi^2 )^{1/3}.
\end{equation}
Representing the function $f_{L,\mu}(\tau)$ as
\begin{equation}
f_{L,\mu}(\tau) = \int \dfrac{d \omega}{2\pi}
f_{L,\mu}(\omega) e^{-i\mu \omega \tau}
\label{fmuf}
\end{equation}
and substituting Eq.~\eqref{fmuf} into the equation for $F_{L,\mu}$, we obtain
\begin{multline}
\Biggl[\mu +
\mu \dfrac{2\sqrt{D}\dot \chi}{(2D |\Delta| \dot \chi^2 )^{1/3}}
\dfrac{d}{d\omega}
\\+
a[\mu|\Delta| + \mu\omega(2D |\Delta| \dot \chi^2 )^{1/3}/4]\Biggr]
f_{L,\mu}(\omega)=0.
\end{multline}
At this point we again make use of the expansion \eqref{aexp} of the function $a(\varepsilon)$ near $\varepsilon=\mu|\Delta|$. Within this approximation, the above equation reduces to
\begin{equation}
\left[
\dfrac{d}{d\omega}
+
\dfrac{1}{2}
\sqrt{\omega}\right]
f_{L,\mu}(\omega)=0,
\end{equation}
which yields
\begin{equation}
f_{L,\mu}(\omega) =
\dfrac{\sqrt{\pi}}{2}
\exp\left(
-
\dfrac{1}{3}
\omega^{3/2}
\right),
\label{fmusol}
\end{equation}
where we fixed the (in general arbitrary) prefactor for our later convenience and the branch of the function $\omega^{3/2}$ is chosen in accordance with the convention introduced after Eq.~\eqref{fmusol+}

The function $f_{L,\mu}(\tau)$ can be expressed in terms of the Airy functions. Performing a sequence of contour deformations and integration variable substitutions, we obtain
\begin{multline}
f_{L,\mu}(\tau) =
\dfrac{\sqrt{\pi}}{2}
\int_{-\infty}^{\infty} \dfrac{d \omega}{2\pi}
\exp\left(
-
\dfrac{1}{3}
\omega^{3/2}
\right) e^{-i\mu\omega \tau}
\\=
- e^{-i\mu\pi/6}
\dfrac{\sqrt{\pi}}{2}
\dfrac{1}{\tau}
\dfrac{\partial}{ \partial  \tau}
\left[
\exp\left(i \mu \dfrac{2}{3} \tau^3\right)
\Ai
\left(\tau^2 e^{i\pi \mu/3}\right)
\right],
\end{multline}
which yields
\begin{multline}
f_{L,\mu}(\tau)
=
- e^{i\mu\pi/6} \sqrt{\pi}
\exp\left(i \mu \dfrac{2}{3} \tau^3\right)
\Bigl[
e^{i\pi \mu/6}  \tau \Ai \left(\tau^2 e^{i\pi \mu/3}\right)
\\+
\Ai' \left(\tau^2 e^{i\pi \mu/3}\right)
\Bigr],
\quad
f_{L,\mu}(\tau) =  f^*_{L,-\mu}(\tau).
\label{fLAiry}
\end{multline}

Employing the asymptotics of the Airy function
\begin{equation}
\Ai(z) =
\dfrac{1}{2\sqrt{\pi}}
\exp\left(-
\dfrac{2}{3}z^{3/2}
\right)
\dfrac{1}{z^{1/4}}
\left[
1 - \dfrac{5}{48} \dfrac{1}{z^{3/2}}
+
\cdots
\right],
\end{equation}
we recover the asymptotic behavior of $f_{L,\mu}(\tau)$
\begin{equation}
f_{L,\mu}(\tau)=
e^{i\pi\mu/4}
\begin{cases}
e^{4i\mu \tau^3/3}
\left[
\sqrt{|\tau|} - \dfrac{i\mu}{8} \dfrac{1}{|\tau|^{5/2}}
\right] , & \tau \to -\infty,
\\
- \dfrac{i\mu}{8}\dfrac{1}{\tau^{5/2}}, & \tau \to \infty,
\end{cases}
\end{equation}

Let us now match the solution formulated in terms of the Airy functions to the adiabatic wave functions evaluated above in Eq. \eqref{adiabatic}. In the vicinity $t \sim t_{2\pi,k,c}$ the adiabatic wave functions take the form
\begin{widetext}
\begin{gather}
\psi_{k, c; L,+}(t) =
-i
\dfrac{
\begin{pmatrix}
\sqrt{R} \\ 1 + c\sqrt{D}
\end{pmatrix}
}{
\sqrt{2}\sqrt{1 + c\sqrt{D}}}
\exp\left(
-i \int_{t_{\pi, k, c}}^{t_{2\pi, k, c}}\varepsilon_A[\chi(\tilde t)] d \tilde t
\right)
\exp\left(
-i|\Delta|(t-t_{2\pi, k, c}) + i \dfrac{4 \tau^3}{3}
\right),
\\
\psi_{k, c; L, -}(t) =
\dfrac{
\begin{pmatrix}
1 + c\sqrt{D} \\ -\sqrt{R}
\end{pmatrix}
}{
\sqrt{2}\sqrt{1 + c\sqrt{D}}}
\exp\left(
i \int_{t_{\pi, k, c}}^{t_{2\pi, k, c}}\varepsilon_A[\chi(\tilde t)] d \tilde t
\right)
\exp\left(
i|\Delta|(t-t_{2\pi, k, c}) - i \dfrac{4 \tau^3}{3}
\right).
\end{gather}
Comparing these expressions with the Airy-function solution, we arrive at the following correspondence
\begin{multline}
\sqrt{|\sin[\chi(t)/2]|} \psi_{k, c;L, +}(t)
\Leftrightarrow
-i
\dfrac{
\begin{pmatrix}
\sqrt{R} \\ 1 + c\sqrt{D}
\end{pmatrix}
}{
\sqrt{2}\sqrt{1 + c\sqrt{D}}}
\sqrt{2}
\left(
\dfrac{\dot \chi}{2D|\Delta|}
\right)^{1/6}
e^{-i\pi/4}
f_+(\tau)
\times\\\times
\exp\left(
-i \int_{t_{\pi, k, c}}^{t_{2\pi, k, c}}\varepsilon_A[\chi(\tilde t)] d \tilde t
\right)
\exp\left[
-i|\Delta|(t-t_{2\pi, k, c})\right],
\label{tpsip}
\end{multline}
and
\begin{multline}
\sqrt{|\sin[\chi(t)/2]|} \psi_{k, c;L,-}(t)
\Leftrightarrow
\dfrac{
\begin{pmatrix}
1 + c\sqrt{D} \\ -\sqrt{R}
\end{pmatrix}
}{
\sqrt{2}\sqrt{1 + c\sqrt{D}}}
\sqrt{2}
\left(
\dfrac{\dot \chi}{2D|\Delta|}
\right)^{1/6}
e^{i\pi/4}
f_-(\tau)
\times\\\times
\exp\left(
i \int_{t_{\pi, k, c}}^{t_{2\pi, k, c}}\varepsilon_A[\chi(\tilde t)] d \tilde t
\right)
\exp\left[
i|\Delta|(t-t_{2\pi, k, c})\right],
\label{tpsim}
\end{multline}
where both sides have identical asymptotic behavior in the region
\begin{equation}
t < t_{2\pi, k, c},
\quad
(2D |\Delta| \dot \chi^2 )^{1/3} |t - t_{2\pi, k, c}| \gg 1,
\quad
|\chi(t) - \chi(t_{2\pi, k, c})| \ll 1.
\end{equation}

\end{widetext}

The matching conditions in Eqs.~\eqref{tpsip} and \eqref{tpsim} imply that the wave functions in the right-hand side should be employed instead of the adiabatic wave functions in the left-hand side provided the time $t$ is in the vicinity of $t_{2\pi,k,c}$. 

\subsection{Both $t$ and $t'$ are close to $t_{0,k,c}$}

Now let us turn to the situation for which both time arguments $t$ and $t'$ in the integral kernel of the inverse operator $(W + a)^{-1}$ remain close to $t_{0,k,c}$. In this regime the kernel has no longer a separable form and depends on $t$ and $t'$ in a nontrivial way. In order to determine this dependence it is necessary to solve an inhomogeneous integral equation
\begin{equation}
(W_{c} + a) F(t,t') = \delta(t - t').
\end{equation}
Making use of the expansion of the function $W_{c}[\chi(t)]$ \eqref{W-ct0} near the point $t_{0,k,c}$, we obtain
\begin{multline}
\left[
\begin{pmatrix}
c\sqrt{D} & \sqrt{R} \\
\sqrt{R} & -c\sqrt{D}
\end{pmatrix}
-
\dfrac{i}{2}\sqrt{D}\dot \chi (t - t_{0, k, c})
+
a
\right]
\\\times
F(t,t') = \delta(t - t').
\label{Ftt}
\end{multline}
The solution of this equation can be expressed as
\begin{multline}
F(t,t') =
\sum_{\mu=\pm}
F_{\mu}(t,t')
\begin{pmatrix}
\sqrt{R}  \\ \mu - c \sqrt{D}
\end{pmatrix}
\mu \dfrac{\mu  +  c \sqrt{D}}{2R}
\\\times
\begin{pmatrix}
\sqrt{R}  & \mu - c \sqrt{D}
\end{pmatrix}.
\label{F1}
\end{multline}
As a result, we obtain the scalar equations for $F_{\mu}(t,t')$:
\begin{equation}
\left[
\mu -
\dfrac{i}{2}\sqrt{D}\dot \chi (t - t_{0, k, c})
+
a\right] F_{\mu}(t,t') = \delta(t - t').
\end{equation}
Similarly to our previous analysis, it would be convenient to factor out the rapidly oscillating phase 
\begin{equation}
F_{\mu}(t,t') = e^{-i\mu|\Delta|(t-t')} f_{\mu}(\tau, \tau'),
\label{Ff}
\end{equation}
where $\tau$ and $\tau'$ are defined by the equations
\begin{gather}
t - t_{0, k, c} = 4 \tau /(2D |\Delta| \dot \chi^2 )^{1/3},
\\
t' - t_{0, k, c} = 4 \tau' /(2D |\Delta| \dot \chi^2 )^{1/3}.
\end{gather}
Representing $f_{\mu}(\tau,\tau')$ in the form
\begin{equation}
f_{\mu}(\tau,\tau') = \int \dfrac{d \omega}{2\pi}
f_{\mu}(\omega,\tau') e^{-i\mu \omega \tau},
\end{equation}
we arrive at the equation 
\begin{equation}
\left[
-\dfrac{d}{d\omega}
+
\dfrac{1}{2}\sqrt{\omega}\right]
f_{\mu}(\omega,\tau')
=
\dfrac{\mu\sqrt{|\Delta|}}{4 \sqrt{2}}
(2D |\Delta| \dot \chi^2 )^{1/6}
e^{i\mu\omega  \tau'},
\end{equation}
the solution of which can be written as
\begin{multline}
f_{\mu}(\tau,\tau')
=
\dfrac{\mu\sqrt{|\Delta|}}{4 \sqrt{2}}
(2D |\Delta| \dot \chi^2 )^{1/6}
\\\times
\int_{-\infty}^{\infty}
\dfrac{d \omega}{2\pi}
\int_{\omega}^{\infty}
d w
e^{\omega^{3/2}/3 - w^{3/2}/3}
e^{-i\mu \omega \tau + i \mu w \tau'},
\label{ftt}
\end{multline}
where the branches of the functions $\omega^{3/2}$ and $w^{3/2}$ are chosen according to the convention introduced after Eq. \eqref{fmusol+}. One readily verifies that $f_{\mu}(\tau,\tau')$ has the retarded structure, i.e. it is proportional to $\theta(\tau-\tau')$. Furthermore, in the limit $\tau,\tau'\to\infty$, the integral kernel \eqref{F1}, with $F$ defined by Eqs. \eqref{Ff} and \eqref{ftt}, reproduces the corresponding asymptotic behavior described by Eq. \eqref{invW3}.

Now we can evaluate the operator $F(h - a h a^+)F^+$. We obtain
\begin{multline}
F(h - a h a^+)F^+
=
\sum_{\mu=\pm}
F_{\mu} (h - a h a^+)
F^+_{\mu}
\\\times
\begin{pmatrix}
\sqrt{R}  \\ \mu - c \sqrt{D}
\end{pmatrix}
\mu \dfrac{\mu  +  c \sqrt{D}}{2R}
\begin{pmatrix}
\sqrt{R}  & \mu - c \sqrt{D}
\end{pmatrix}
\end{multline}
and
\begin{multline}
F_{\mu} (h - a h a^+) F^+_{\mu}
=
h_{\mu}
\dfrac{\sqrt{|\Delta|}}{4\sqrt{2}}
(2D |\Delta| \dot \chi^2 )^{1/6}
e^{-i\mu|\Delta|(t-t')}
\\\times
2\pi
\int_{-\infty}^{\infty}
\dfrac{d \omega}{2\pi}
\int_{-\infty}^{\infty}
\dfrac{d \omega_1}{2\pi}
e^{\omega^{3/2}/3 + \omega_1^{3/2}/3 - 2[\max(0,\omega,\omega_1)]^{3/2}/3}
\\\times
e^{-i\mu \omega \tau + i\mu \omega_1 \tau'}.
\end{multline}

In order to evaluate the corresponding contribution to the current one needs to employ the combination
\begin{equation}
W^+_{c}(\chi) \partial_{\chi} W_{c}(\chi)
=
-
\dfrac{i c  \sqrt{D}}{2}
\begin{pmatrix}
\sqrt{D} & \sqrt{R} e^{i c \chi/2} \\
\sqrt{R} e^{-i c \chi/2}  & -\sqrt{D}
\end{pmatrix},
\end{equation}
which reduces to
\begin{equation}
W^+_{c}(\chi) \partial_{\chi} W_{c}(\chi)
=
-
\dfrac{i c  \sqrt{D}}{2}
\begin{pmatrix}
\sqrt{D} & c \sqrt{R} \\
c \sqrt{R}  & -\sqrt{D}
\end{pmatrix}
\end{equation}
at $t = t_{0,k,c}$. We then obtain
\begin{multline}
i e
\Sp \left\{F(h - a h a^+)F^+ W^+_{c}(\chi) \partial_{\chi} W_{c}(\chi)\right\}
\\=
e
\dfrac{\sqrt{D|\Delta|}}{8\sqrt{2}}
(2D |\Delta| \dot \chi^2 )^{1/6}
\sum_{\mu}
\mu
h_{\mu}
e^{-i\mu|\Delta|(t-t')}
\\\times
2\pi
\int_{-\infty}^{\infty}
\dfrac{d \omega}{2\pi}
\int_{-\infty}^{\infty}
\dfrac{d \omega_1}{2\pi}
e^{\omega^{3/2}/3 + \omega_1^{3/2}/3 - 2[\max(0,\omega,\omega_1)]^{3/2}/3}
\\\times
e^{-i\mu \omega \tau + i\mu \omega_1 \tau'}.
\label{Idisstau}
\end{multline}
By setting $\tau=\tau'$ in Eq. ~\eqref{Idisstau} one could evaluate the corresponding contribution to the current originating from the vicinity of the point $t_{0,k,c}$. It turns out, however, that this limit is singular due to the divergence of the integral \eqref{Idisstau} at  $\tau\to\tau'$. In order to cure this divergence it is necessary to also evaluate the analogous contribution arising from the  endpoint $t_{2\pi,k+c/2-1/2,-c}$ from the neighboring time interval. Since $t_{2\pi,k+c/2-1/2,-c}=t_{0,k,c}$ this contribution affects the current in the same time region. As we will see, the latter contribution also contains a similar divergence. Below we will demonstrate that the sum of these two contributions becomes regular since the corresponding singularities cancel each other in the limit $\tau\to \tau'$.

\subsection{Both $t$ and $t'$ are close to $t_{2\pi,k,c}$}

Provided both time arguments $t$ and $t'$ in the kernel of the operator $(W + a)^{-1}$ are close to $t_{2\pi,k,c}$, the whole calculation is similar to the one carried out above.  We now need to solve the equation
\begin{equation}
(W_{c} + a) F(t,t') = \delta(t - t')
\end{equation}
which at $t$ and $t'$ near $t_{2\pi,k,c}$ reduces to the form
\begin{multline}
\left[
\begin{pmatrix}
-c\sqrt{D} & \sqrt{R} \\
\sqrt{R} & c\sqrt{D}
\end{pmatrix}
+
\dfrac{i}{2}\sqrt{D}\dot \chi (t - t_{2\pi, k, c})
+
a
\right]
\\\times
F(t,t') = \delta(t - t')
\label{Ftt2}
\end{multline}
with the solution 
\begin{multline}
F(t,t') =
\sum_{\mu=\pm}
F_{\mu}(t,t')
\begin{pmatrix}
\sqrt{R}  \\ \mu + c \sqrt{D}
\end{pmatrix}
\mu \dfrac{\mu  -  c \sqrt{D}}{2R}
\\\times
\begin{pmatrix}
\sqrt{R}  & \mu + c \sqrt{D}
\end{pmatrix},
\end{multline}
where the scalar functions $F_{\mu}(t,t')$ satisfy
\begin{equation}
\left[
\mu +
\dfrac{i}{2}\sqrt{D}\dot \chi (t - t_{2\pi, k, c})
+
a\right] F_{\mu}(t,t') = \delta(t - t').
\end{equation}

Proceeding in much the same way as above, we eventually arrive at the contribution to the current in the form
\begin{multline}
i e
\Sp \left\{F(h - a h a^+)F^+ W^+_{c}(\chi) \partial_{\chi} W_{c}(\chi)\right\}
\\=
-e
\dfrac{\sqrt{D|\Delta|}}{8\sqrt{2}}
(2D |\Delta| \dot \chi^2 )^{1/6}
\sum_{\mu}
\mu
h_{\mu}
e^{-i\mu|\Delta|(t-t')}
\\\times
2\pi
\int_{0}^{\infty}
\dfrac{d \omega}{2\pi}
\int_{0}^{\infty}
\dfrac{d \omega_1}{2\pi}
e^{-\omega^{3/2}/3 - \omega_1^{3/2}/3}
\Bigl[
e^{2[\min(\omega,\omega_1)]^{3/2}/3} -1
\Bigr]
\\\times
e^{-i\mu \omega \tau + i\mu \omega_1 \tau'}.
\label{Idisstau2}
\end{multline}

\subsection{The current}

We now combine Eqs.~\eqref{Idisstau} and \eqref{Idisstau2}. For $\chi \simeq 2\pi n$, the current is determined by contributions from two time intervals: $t,t' \approx t_{0,k,c}$ within the interval  $(k,c)$  and $t,t' \approx t_{2\pi,k+c/2-1/2,-c}$ within the neighboring interval $(k+c/2-1/2,-c)$.  Combining these terms and taking the equal time limit, we observe that all singular contributions cancel and we obtain
\begin{equation}
I_{23}(t) =
e D |\Delta|
\left(
\dfrac{\dot \chi}{2D|\Delta|}
\right)^{1/3}
\sum_{\mu}
\mu
h_{\mu}
|f_{L,\mu}(-\tau)|^2.
\label{I23}
\end{equation}

Making use of the identity
\begin{equation}
|f_{L,\mu}(-\tau)|^2 = |f_{L,\mu}(\tau)|^2 + \tau,
\end{equation}
and combining Eqs.~\eqref{I3} and \eqref{I23}, we recover the total current
\begin{multline}
I(t)=I_1(t)+I_{23}(t)=
e D |\Delta|
\left(
\dfrac{\dot \chi}{2D|\Delta|}
\right)^{1/3}
[h_{+} - h_{-}]
\tau
\\+
2 e D |\Delta|
\left(
\dfrac{\dot \chi}{2D|\Delta|}
\right)^{1/3}
[h_{+} - h_{-}]
|d(t)|^2
|f_{L,+}(\tau)|^2.
\label{123}
\end{multline}

\end{document}